# A Kinetic Model for Cell Damage Caused by Oligomer Formation

Liu Hong*, Ya-Jing Huang, Wen-An Yong

Zhou-Pei Yuan Center for Applied Mathematics, Tsinghua University, Peking, 100084, P.R. China

*zcamhl@tsinghua.edu.cn

**Abstract:** It is well-known that the formation of amyloid fiber may cause invertible damage to cells, while the underlying mechanism has not been fully uncovered. In this paper, we construct a mathematical model, consisting of infinite ODEs in the form of mass-action equations together with two reaction-convection PDEs, and then simplify it to a system of 5 ODEs by using the maximum entropy principle. This model is based on four simple assumptions, one of which is that cell damage is raised by oligomers rather than mature fibrils. With the simplified model, the effects of nucleation and elongation, fragmentation, protein and seeds concentrations on amyloid formation and cell damage are extensively explored and compared with experiments. We hope that our results can provide a valuable insight into the processes of amyloid formation and cell damage thus raised.

## Introduction

Since the first discovery of prions by Prusiner in 1982 [Prusiner82], more than twenty different kinds of human neuron-degenerative diseases, such as Alzheimer's and Parkinson's diseases, have been identified to be correlated with the abnormal accumulation of amyloid proteins and fibrils in tissues [Chiti06]. As a result, to quantify the relationship between the processes of amyloid formation and cell damage thus caused has become a central task in this field.

As a typical self-assembling bio-system, the study of amyloid fiber has a long history and has attracted broad interests. Pioneer works can be dated back to Oosawa for his studies on actins in the late 1950s [Oosawa59]. Later, the roles of conformational transition, primary nucleation and elongation during the formation of amyloid formation were gradually explored and explained [Morris09]. Hofrichter, Eaton and Ferrone found that surface-catalyzed secondary nucleation is crucial for several amyloid proteins, *e.g.* sickle-cell hemoglobin and $IAPP_{20-29}$ [Hofrichter74, Ferrone80, Ruschak07]. Although fragmentation as an alternative way for secondary nucleation was already mentioned in Oosawa's famous book [Oosawa75], it is the beginning of new era when corresponding detailed experimental validations [Collions04, Xue09] and mathematical treatments [Knowles09, Hong13] have been performed. Now we are at a stage to interpret and predict various kinetic and thermodynamic data in real-time with high precision [Cohen12, Tan13]. This provides a good foundation for our current study.

For quite a long time, the mature fiber has been widely taken as the major cause of cell damage, and the fiber concentration in tissues was regarded as a key factor to quantify the progress of amyloidosis [Dobson03, Knowles09]. The more fibrils deposit, the greater toxicity will be raised to cells. However with accumulative evidences on the morphology, structure and function of oligomers and fibrils in vitro and in vivo [Haass07, Simone08], this view has been largely changed in the last several years. An alternative possible mechanism was proposed that oligomers may bind onto the cell membrane, which can dramatically change the local environment and geometry of the membrane, and even cause the rupture of the latter. These changes will give raise to abnormal ion leakage, especially $Ca^{2+}$ flux, which is fatal to cells [Willams11, Wong09].

Despite extensive studies on the formation of amyloid fiber, relatively limited researches have been carried on oligomers [Ono09, Fandrich12]. In this paper, we attempt to build a mathematical model which can provide a general physical picture on the processes of amyloid formation and the corresponding cell damage caused by oligomer-cell interaction together. Our model is based on four basic assumptions, and formulated through a group of coupled mass-action equations and reaction-convection equations. By using the method of maximum entropy principle, the derived model is greatly simplified into five moment-closure equations, which are ODEs and easy to be solved. Detailed model analyses and comparisons with experimental data are also highlighted.

The current paper is organized as follows: Firstly, we introduce the four basic assumptions adopted. Later, a general framework of our mathematical model and the simplified moment-closure equations are described. Further numerical verifications are shown in Fig. 2. The original mass-action equations plus two reaction-convection equations and the method we used for deriving moment-closure equations (maximum entropy principle) are left to Supporting Materials. Interested readers may find all necessary details there. In the following, our main results are separated into several parts. With the constructed model, the effects of nucleation and elongation, fragmentation, protein and seeds concentrations on amyloid formation and cell damage are extensively explored and compared with experiments. Besides, several interesting scaling laws for characterizing kinetic quantities are mentioned too. Finally, some conclusions and further discussions are addressed.

**The processes of amyloid formation and cell damage are closely correlated, during which oligomers play a key role.**

It is widely known that the processes of amyloid formation and cell damage are closely correlated, but the underlying quantitative relations have never been clearly clarified. To solve this problem, we construct a general mathematical model, which relies on the following four basic assumptions:

i) The basic procedure of amyloid formation can be well characterized through a kinetic model, which includes primary nucleation, elongation and fragmentation, as well as their corresponding inverse processes;

ii) Cell toxicity is mainly caused by oligomers, rather than mature fibrils and monomers, through their binding to cell membrane. The more oligomers have bound, the greater damage to the cell membrane will occur;

iii) The ion concentration inside a cell can be used as an effective index to quantify the cell toxicity. Besides normal ion exchanges, the appearance of additional abnormal ion leakage depends on the degrees of how cell membrane is damaged;

iv) Oligomer binding does not apparently influence the kinetics of amyloid formation. In other words, we neglect the consumption of oligomers during their interactions with the cell membrane.

The first assumption points out the possible mechanisms for amyloid formation. Based on our own studies and other related works [Knowles09, Hong13], most experiments and numerical simulations on amyloid formation can be well interpreted by this kind of model. The necessities of conformational transition [Serio00, Linden07], off-pathway polymerization [Powers08], surface catalysed secondary nucleation [Cohen13], thickening [Pallitto01, Mauro07], protein diffusion [Schreck13] and so on will be left to future studies. The next two assumptions point out possible candidates and mechanisms for cytotoxicity. Although at a first glance, amyloidosis is easy to be linked with the deposit of fibrils in tissues, accumulative evidences in recent years gradually change this view and reveal that oligomers may be the prime pathogenic factor instead. The cytotoxicity of monomers and fibrils are often negligible compared to that of oligomers [Doran12, Lorenzen14]. Here we make a further assumption on the possible mechanism of cytotoxicity, including oligomer binding, cell membrane damage and abnormal ion leakage. The final degrees of cytotoxicity are quantified through the leaked ion concentration compared with its normal values. The last one hints that the processes of amyloid formation are independent of oligomer binding. In other words, we neglect the consumption of oligomers during their interactions with the cells. Although it seems to be a very rough approximation, it still works in most cases since the number concentration of oligomers is generally much higher than cells. As we will see, this assumption will help us to largely simplify the mathematical modeling and computation.

**The kinetics of amyloid formation and cell damage both can be well modeled through a series of chemical reactions.**

Based on our assumptions above, two basic procedures should be considered to model the kinetics of cell damage caused by amyloid formation: one is the formation of oligomers and fibrils; the other is the change of ion concentrations inside a cell, which is caused by the damage of cell membrane through oligomer binding.

The modeling of the first procedure has been well established in the past studies [Oosawa75, Knowles09, Hong13]. Primary nucleation, elongation and fragmentation are generally considered as three basic processes in this part. To be exact, we assume that monomers will firstly aggregate into oligomers through primary nucleation with the critical size $n_c$. Then oligomers grow into mature fibrils by elongation. Once oligomers and fibrils exceed a certain given size, they may break into two small pieces spontaneously. To sum up, this part is modeled through following chemical reactions

$$n_c A_1 \underset{k_n^-}{\overset{k_n^+}{\rightleftharpoons}} A_{n_c},$$

$$A_1 + A_i \underset{k_e^-}{\overset{k_e^+}{\rightleftharpoons}} A_{i+1}, \qquad (i \geq n_c) \tag{1}$$

$$A_{i+j} \underset{k_f^-(i,j)}{\overset{k_f^+(i,j)}{\rightleftharpoons}} A_i + A_j, \qquad (i, j \geq n_c)$$

where $k_n^+$ and $k_e^+$ are the forward reaction rate constants for fiber nucleation and elongation respectively; $k_n^-$ and $k_e^-$ are the corresponding backward ones.

$k_f^+(i,j)$ and $k_f^-(i,j)$ are length-dependent reaction rate constants for fiber fragmentation and association, whose formulas are taken according to the Hill's model [Hill83]. $A_1$ stands for monomers, $A_i$ for oligomers or fibrils of size $i$. Here we distinguish oligomers and fibrils based on their respective sizes and functions rather than the morphology and reaction rates. Oligomers are supposed to be small aggregates with size varying from $n_c$ to $n_o$ and can cause cytotoxicity; while fibrils has size larger than $n_o$ and thus do not cause cell damage. Above descriptions actually provide a practical (rather than precise) definition of oligomers and fibrils. And its benefits will be clearly seen through the modeling in the next part.

In the second part, we need to model the interaction between oligomers and cells. As many recent literatures have revealed [Fandrich12, Stefani13], the binding of oligomers onto the cell membrane may cause the breakage of the latter, give rise to the leakage of ions (especially Ca$^{2+}$) and eventually the death of cells. Based on this general picture, we assume that the cells under consideration can adopt two different states: one is the normal state (denoted by $C$), the other is the damaged state caused by oligomer binding (denoted by $C^*$). These two states can converge into each other under certain conditions. The forward process (from $C$ to $C^*$) depends on the oligomer concentration, while the backward one (from $C^*$ to $C$) does not (very weak and slow self-healing of the membrane), whose reaction rate constants are denoted as $k_b^+$ and $k_b^-$ respectively.

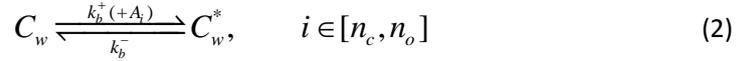

$$C_w \underset{k_b^-}{\overset{k_b^+(+A_i)}{\rightleftarrows}} C_w^*, \qquad i \in [n_c, n_o] \tag{2}$$

Here a lower index $w \in [0,1]$ is added to denote the normalized ion concentration inside a cell (therefore we classify the cells not only by whether they are damaged or not, but also by the ion concentration inside). Its value can be taken as an effective quantity to measure the cell damage raised by amyloid formation and is directly related to most popular experiments on cytotoxicity.

The changes of ion concentrations inside a cell can be attributed to two basic processes: normal ion exchange and abnormal ion leakage due to the damaged cell membrane. Here for simplicity, we assume ion exchange only exists for normal cells while ion leakage for damaged cells. This point will bring no trouble for modeling even if both processes are considered for both types of cells. Consequently, we have

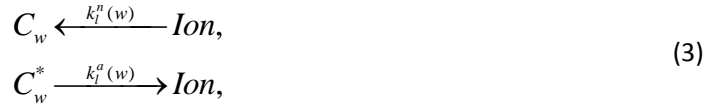

$$\begin{aligned} C_w &\xleftarrow{k_l^n(w)} Ion, \\ C_w^* &\xrightarrow{k_l^a(w)} Ion, \end{aligned} \tag{3}$$

where $k_l^n(w)$ is the reaction rate constant for normal ion exchange and $k_l^a(w)$ for the abnormal ion leakage. According to famous Fick's law, we may take $k_l^n(w) = 2k_l^n \cdot (1-w)$ and $k_l^a(w) = 2k_l^a \cdot w$ in general.

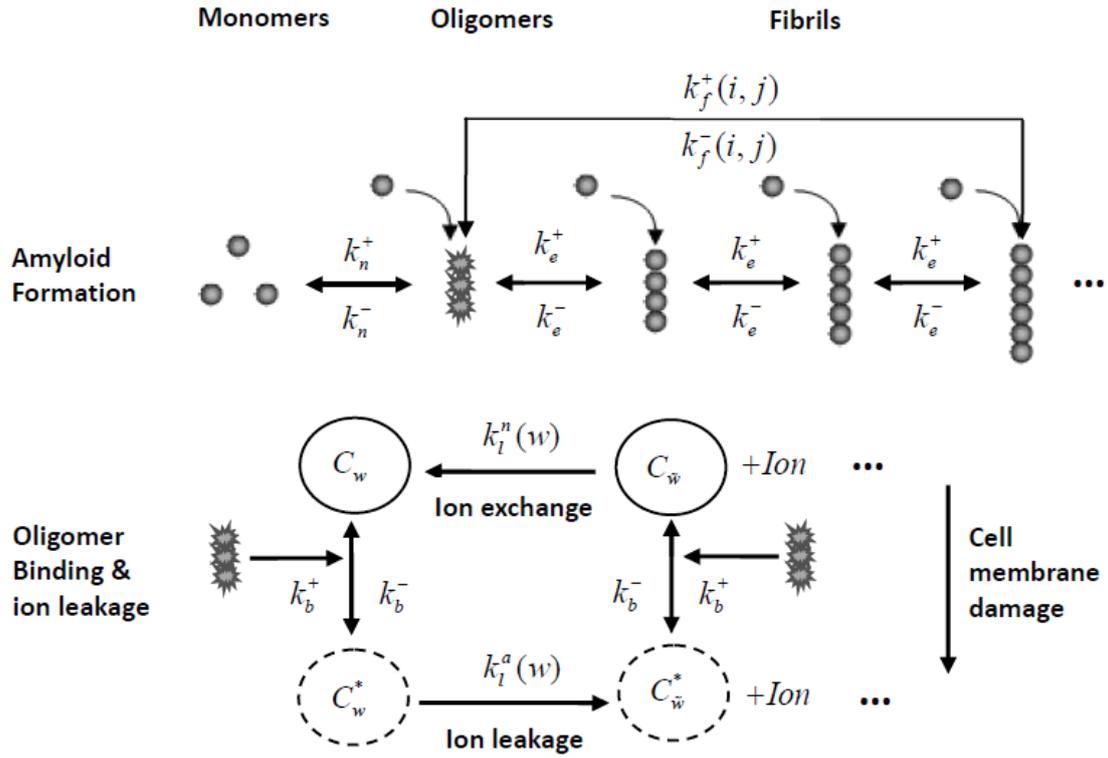

Fig 1. An illustration of our kinetic model with oligomer size as 3 and $w > \tilde{w}$.

**Moment-closure equations provide an efficient way to quantify amyloid formation and cell damage in replace of mass-action equations.**

With the chemical reactions listed in Eq. 1 in hand, it is easy to formulate them into a system of ordinary differential equations according to the laws of mass action. While Eqs. 2 and 3 together represent two additional partial differential equations for oligomer binding and ion leakage. These equations together constitute the basic model of our current study (Eq. S1). However, instead of bothering with these complicated formulas, we will turn to simple moment-closure equations. Interested readers may refer to Supporting Materials for details on how to formulate mass-action equations and reaction-convection equations, how to derive moment-closure equations based on the method of maximum entropy principle, and so on.

Following the moment-closure method introduced in our previous studies [Hong13, Tan13], two macroscopic moments, namely the number concentration of aggregates (including both oligomers and fibrils) $P = \sum_{i=n_c}^{\infty} [A_i]$ and the mass concentration of aggregates $M = \sum_{i=n_c}^{\infty} i \cdot [A_i]$ are adopted to characterize the kinetics of amyloid formation, where $[A_i]$ stands for the concentration of aggregates of size $i$. Besides them, four additional quantities are introduced for cell damage and ion leakage in the same way, i.e. $C_+ = \int_0^1 [C(w,t) + C^*(w,t)] dw$, $C_- = \int_0^1 [C(w,t) - C^*(w,t)] dw$, $I_+ = \int_0^1 w[C(w,t) + C^*(w,t)] dw$, and $I_- = \int_0^1 w[C(w,t) - C^*(w,t)] dw$, where $C(w,t)$ and $C^*(w,t)$ stand for the concentration of normal and damaged cells with ion concentration $w$ at time $t$ respectively. It is easily noticed that $C_+(t) = c_{tot}$ is a constant representing the conservation of total cells.

Besides $C_+$ which is a constant, a self-closed system of ODEs for the other five quantities can be derived from original coupled mass-action equations and reaction-convection equations (Eq. S1) by applying the moment-closure method given in Supporting Materials. In this way, we are able to reduce the dimension of our mathematical model from infinite (actually infinite-dimensional ODEs plus two PDEs) to 5 (two for amyloid formation plus three for cell damage and ion leakage)

$$\begin{cases} \dfrac{d}{dt}P = k_n^+(m_{tot}-M)^{n_c} - k_n^-(1-\theta)P + \sum_{i=n_c}^{\infty}\sum_{j=i+n_c}^{\infty} k_f^+(i,j-i)(1-\theta)\theta^{j-n_c}P \\ \qquad - \sum_{i=n_c}^{\infty}\sum_{j=n_c}^{\infty} k_f^-(i,j)(1-\theta)^2 \theta^{i+j-2n_c}P^2, \\ \dfrac{d}{dt}M = n_c k_n^+(m_{tot}-M)^{n_c} - n_c k_n^-(1-\theta)P + 2k_e^+(m_{tot}-M)P - 2k_e^-\theta P, \\ \dfrac{d}{dt}C_- = -k_b^+ P_{oli}(c_{tot}+C_-) + k_b^-(c_{tot}-C_-), \\ \dfrac{d}{dt}I_+ = k_l^n(c_{tot}+C_-) - k_l^n(I_+ + I_-) - k_l^a(I_+ - I_-), \\ \dfrac{d}{dt}I_- = k_l^n(c_{tot}+C_-) - (k_l^n + k_b^+ P_{oli})(I_+ + I_-) + (k_l^a + k_b^-)(I_+ - I_-), \end{cases} \quad (4)$$

where $\theta \equiv (M - n_c P)/[M - (n_c - 1)P] \in [0,1)$. $P_{oli} \equiv \sum_{j=n_c}^{n_o}[A_j] = \sum_{j=n_c}^{n_o}(1-\theta)\theta^{j-n_c}P$ and $M_{oli} \equiv \sum_{j=n_c}^{n_o} j \cdot [A_j] = \sum_{j=n_c}^{n_o} j \cdot (1-\theta)\theta^{j-n_c}P$ denote the number and mass concentrations of oligomers respectively. The initial conditions can be generally taken as $P(0) = M(0) = 0$ and $C_-(0) = I_+(0) = I_-(0) = c_{tot}$ in the absence of initial seeds and damaged cells. It is noted that the benefits of our fourth assumption can be clearly seen from above formulas, which in fact allows us to decouple the processes of amyloid formation from oligomer binding mathematically. In other words, amyloid formation manipulates the degrees of cell damage by controlling the concentration of oligomers, while oligomer binding gives no feedback effect.

Compared to original coupled mass-action equations and reaction-convection equations (Eq. S1), our new moment-closure equations are tremendously simple. And the computational efficiency has been improved by at least 10, 000 times (from several days to seconds). Therefore Eq. 4 is very suitable for the real-time analysis of various experimental data on amyloid formation and correlated cell damage.

**The kinetics of amyloid formation and cell damage are well presented by the moment-closure equations.**

A major difference between our model and previous ones lies on that there is no apparent correlation between the processes of fiber generation and cell damage. In previous studies, as mature fibrils are assumed to be responsible for cytotoxicity, the processes of fiber generation and cell damage are positively correlated. The more fibrils are formed, the more serious cell damage will be caused. Contrarily in our current model, according to the second assumption that monomers and mature fibrils are unharmful to cells, the degree of cell damage does not depend on the fiber concentration. Furthermore, as the oligomer concentration stays very low during the whole process of amyloid formation (comparing to that of monomers and fibrils), the progress of cell damage will exhibit a much complicated dependence on the kinetic details of amyloid formation as expected.

Are the kinetic essentials of amyloid formation and cell damage well preserved during our simplification procedure? Can the moment-closure equations honestly reflect the time-evolutionary behaviors of moments which we care about? These questions are crucial for the validity and applicability of our moment-closure equations. In Fig. 2, we carefully compare numerical solutions of our moment-closure equations (Eq. 4) and original coupled mass-action equations and reaction-convection equations (Eq. S1). It is clearly seen from Figs. 2A and 2B that except for the static solutions of $P(t)$, whose difference is mainly caused by the exponential fiber length distribution inappropriately assumed in the moment-closure method (or in other words, the entropy function we used is not so precise), both models predict very similar results on the fibril concentrations. Even for the mass and number concentrations of oligomers, which play a key role in linking the two processes of amyloid formation and cell damage, perfect agreements are observed in the early and late time regions (Fig. 2C). The only difference lies in the middle region -- the exponential growth phase, during which the elongation process quickly promotes most nucleuses into fibrils and thus leads to a dramatic decrease of oligomers. Despite this little disagreement in kinetics, the predictions of the moment-closure equations seem quite satisfactory on the concentration of damage cells and the amount of ion leakage through the membrane (Fig. 2D). These agreements confirm the fact that the kinetic essentials of amyloid formation and cell damage are well preserved during our simplification procedure. And thus the moment-closure equations provide an honest yet much efficient way to study the time-evolutionary behaviors of moments that we care about.

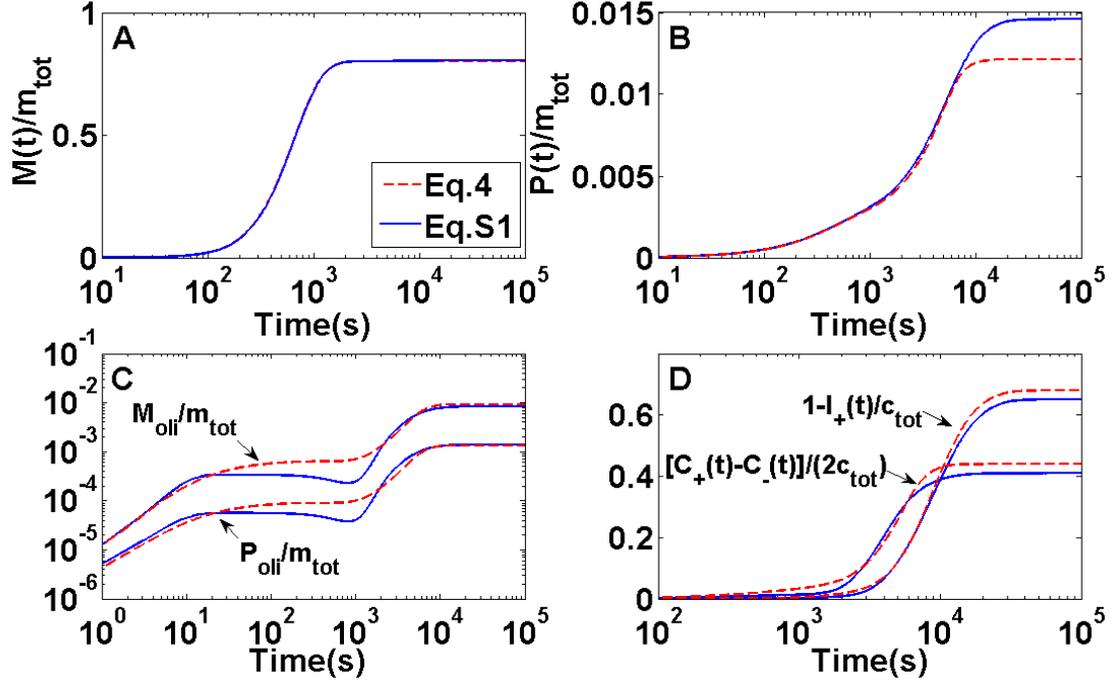

Fig 2. Comparisons of the moment-closure equations and coupled mass-action equations together with reaction-convection equations on amyloid formation and cell damage. (A) The mass concentration of aggregates; (B) the number concentration of aggregates; (C) the number and mass concentrations of oligomers; (D) the degrees of cell damage characterized through the percentage of damaged cells $[C_+(t)-C_-(t)]/(2c_{tot})$ and the amount of ion leakage $1-I_+(t)/c_{tot}$. Blue solid lines stand for solutions of original coupled equations (Eq. S1); while red dashed lines for solutions of moment-closure equations (Eq. 4). Here we set $m_{tot}=5\times10^{-5}M$, $c_{tot}=4.3\times10^{-5}M$, $k_n^+=0.1M^{-1}s^{-1}$, $k_n^-=0.001s^{-1}$, $k_e^+=10^4M^{-1}s^{-1}$, $k_e^-=0.1s^{-1}$, $k_f^+=10^{-4}s^{-1}$, $k_f^-=8\times10^4M^{-1}s^{-1}$, $k_b^+=10^4M^{-1}s^{-1}$, $k_b^-=0.001M^{-1}s^{-1}$, $k_l^n=5\times10^{-5}M^{-1}s^{-1}$, $k_l^a=1.5\times10^{-4}M^{-1}s^{-1}$, $n_c=2$, $n_o=10$ and $n=1$.

**High nucleation rate speeds up cell damage, while high elongation rate generally slows down it.**

Now we plan to use our moment-closure equations to systematically explore the effects of primary nucleation and elongation on the processes of amyloid formation and especially cell damage (see Figs. S1 and S2 in Supporting Materials). It is clearly seen that with the increase of primary nucleation rates, speeds for fibril formation, oligomer formation and ion leakage (or membrane leakage) grow simultaneously [Doran12]. This phenomenon is easy to understand by noticing the fact that primary nucleation is a direct way to produce nucleuses (or oligomers in this case). The effect of elongation rate seems to be much more complex. Generally speaking, the elongation rate is positively correlated with the speed of fibril formation and is inversely proportional to the speed of oligomer formation and cell damage. Such a behavior is mainly due to our second assumption. If cell damage is mainly caused by fibrils, then for sure the degrees of cell damage will be positively proportional to the elongation rate too. However, in our model cell damage is assumed to be raised by oligomers rather than fibrils. As a consequence, high elongation rate would suppress the generation of oligomers and thus the degrees of cell damage in the end. With a further increase in the elongation rate, this suppression effect will become less and less apparent due to the limited number of oligomers [Wang11].

**Fragmentation can dramatically accelerate the formation of amyloid fiber and give rise to more serious damage to cells.**

As a most important way for secondary nucleation, fragmentation plays a key role in the formation of breakable filaments. Once a fibril is breakable, nucleuses (including ogligomers and short fibrils) are easy to be generated from long mature fibrils rather than the classical slow way through primary nucleation. Through such a mechanism, fragmentation can dramatically speed up the formation of fibrils and oligomers, as well as the processes of cell damage (see Figs. S3 in Supporting Materials) [Xue09, Xue10]. However, in the absence of initial nucleuses, the influence of fragmentation is usually not as timely as the primary nucleation, since fragmentation can only be dominant once the fiber concentration exceeds a certain threshold through primary nucleation.

In the last ten years, the effect of fragmentation on amyloid formation and cell damage has been widely explored in experiments. Here as a typical example, we apply our model to analyze the agitation data of $\beta_2 m$ fibrils performed by Xue *et al.* [Xue09]. Since in this case agitation is dominated, we simply neglect the processes of primary nucleation and elongation. In Fig. 3A, it is clearly seen that the average fiber length decreases monotonically during agitation, which is a strong evident for fiber fragmentation. Much detailed information could be obtained through the fiber length distribution as shown in Fig. 3B, in which the initial Gaussian-like distribution turns into an exponential distribution and more and more samples accumulate in the region of short fibrils or oligomers with time evolving. Most importantly, with agitation going on, the thus prepared fibrillar samples become easy to give rise to a high efficiency in the liposome dye release (Fig. 3C). This result is consistent with the general belief that oligomers and short fibrils are much more harmful to cells than monomers and mature long fibrils [Ono09, Fandrich12].

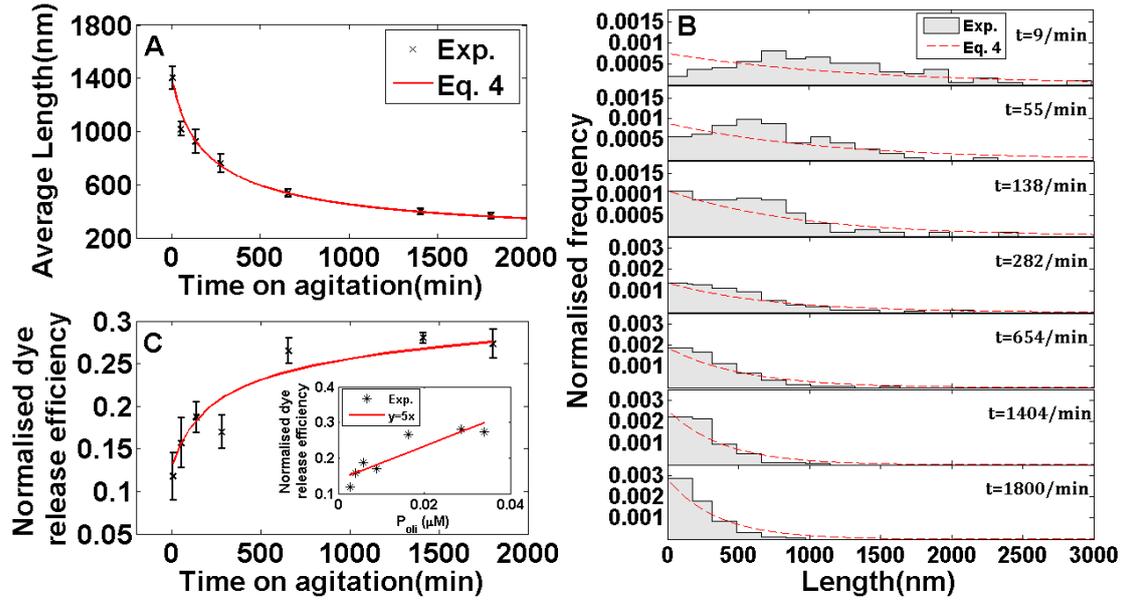

Fig 3. Effects of fragmentation on $\beta_2 m$ fibrils and membrane leakage. (A) The average fiber length under agitation; (B) the fiber length distribution characterized by TM-AFM; (C) the amount of liposome dye release for fibrillar samples under different agitation time. Black stars with error bars stand for the experiment data measured by Xue *et al.* [Xue09]; while red dashed lines for solutions of Eq. 4. Here we choose $m_{tot} = 12 \times 10^{-6} M$, $c_{tot} = 4.3 \times 10^{-5} M$, $k_f^+ = 1 \times 10^{-9} s^{-1}$, $k_f^- = 5 \times 10^3 M^{-1} s^{-1}$, $k_b^+ = 1 \times 10^4 M^{-1} s^{-1}$, $k_b^- = 2.8 \times 10^{-5} M^{-1} s^{-1}$, $k_l^n = 1 \times 10^{-3} M^{-1} s^{-1}$, $k_l^a = 5 \times 10^{-6} M^{-1} s^{-1}$, $n_c = 2$, $n_o = 50$ and $n = 3$. $k_n^+$, $k_n^-$, $k_e^+$ and $k_e^-$ are all set to be zero.

**Protein concentration and seeds concentration have crucial influences on the kinetics of amyloid formation and cell damage.**

It is well-known that the processes of amyloid formation and cell damage sensitively depend on the initial protein concentration and seeds concentration [Bucciantini02]. In the current case, we further address this point by examining the experiments performed by Engel *et al.* [Engel08] on the kinetics of hIAPP fibril growth and hIAPP-induced membrane leakage. In Fig. 4A and 4B, four different cases with initial protein concentration varying from $5\mu M$ to $0.1\mu M$ are studied by the moment-closure equations. Generally speaking, the kinetics of amyloid formation and membrane leakage both sensitively depend on the protein concentration. High initial protein concentration not only means short lag-time and fast fiber growth rate, but also is related to large values of final oligomer and fibril concentrations and membrane leakage in the static state. Although no kinetic trace for the fibril formation is available under three low protein concentrations (except for the half-time given in Fig. 4C), excellent agreements on the kinetics of fibril fromation and normalized membrane leakage still strongly confirm the usefulness and reliability of our model. A further validation on a complete data set of the half-time is made between experimental data and our model predictions in Fig. 4C. Besides the initial protein concentration, the effect of seeding on membrane leakage is explored through Fig. 4D, in which 0%, 1%, 2% and 10% hIAPP seeds have been added independently to the system. It is clearly seen that high concentration of seeds can dramatically speed up the membrane leakage, but have little influence on the final static values as expected.

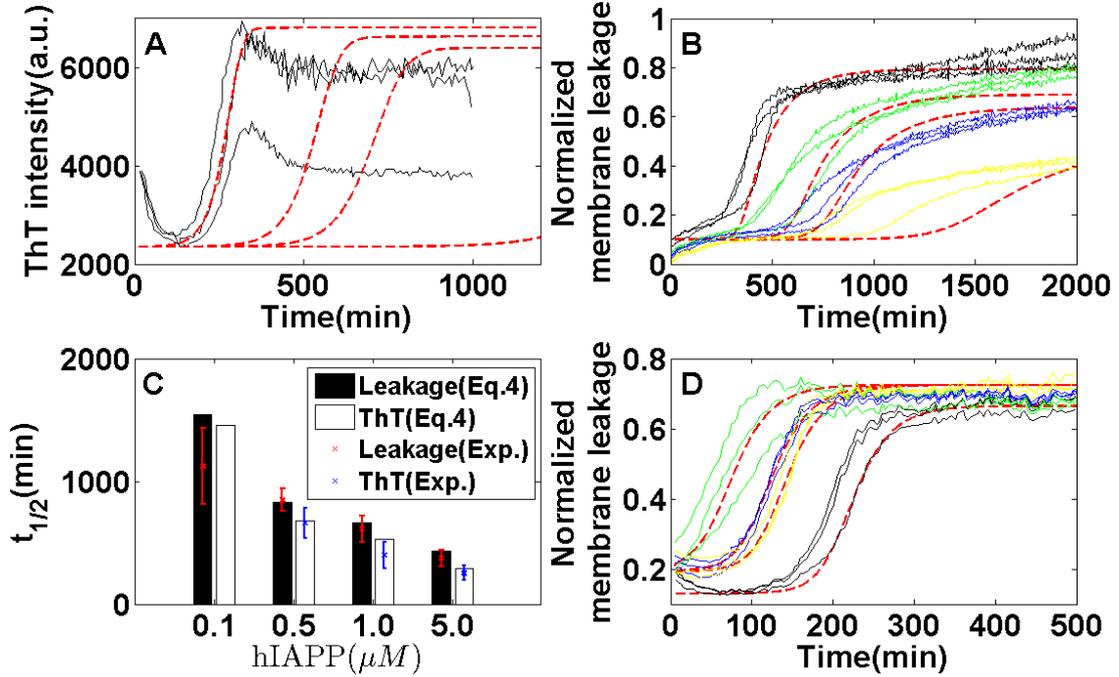

Fig 4. Effects of protein concentrations and seeds concentrations on the kinetics of hIAPP fibril growth and hIAPP-induced membrane leakage. (A) The measured ThT fluorescence intensity; (B) the amount of membrane leakage under different initial protein concentrations, that are $5\mu M$ (black lines), $1\mu M$ (green lines), $0.5\mu M$ (blue lines) and $0.1\mu M$ (yellow lines) respectively; (c) comparisons of the half-time for fibril formation and membrane leakage between our model and experiments; (D) the amount of membrane leakage under different concentrations of seeds, *i.e.* 0% (yellow lines), 1% (blue lines), 2% (green lines) and 10% (black lines) hIAPP seeds respectively. Solid lines stand for experimental data measured by Engel *et al.* [Engel08]; while red dashed lines for solutions of Eq. 4. Here we set $c_{tot}=4.3\times10^{-5}M$, $k_n^+=3\times10^{-5}M^{-1}s^{-1}$, $k_n^-=10^{-4}s^{-1}$, $k_e^+=10^5 M^{-1}s^{-1}$, $k_e^-=5\times10^{-3}s^{-1}$, $k_f^+=7\times10^{-5}s^{-1}$, $k_f^-=10^8 M^{-1}s^{-1}$, $k_b^+=8\times10^4 M^{-1}s^{-1}$, $k_b^-=3.44\times10^{-5}M^{-1}s^{-1}$, $k_l^n=k_l^a=0.05M^{-1}s^{-1}$, $n_c=2$, $n_o=50$ and $n=1$. In subplot (D), $m_{tot}=2.5\times10^{-5}M$, $k_b^+=2\times10^5 M^{-1}s^{-1}$, $k_b^-=2.58\times10^{-4}M^{-1}s^{-1}$ and $M(0)/P(0)=400$.

**Kinetic essentials of amyloid formation and cell damage can be quantitatively learned from simple scaling relations.**

After a qualitative description of the basic kinetics of amyloid formation and cell damage, as well as the influences of primary nucleation and elongation, fragmentation, protein and seeds concentrations, in this section we pursue a quantitative picture. It is already known that the speed of amyloid formation is positively related to the reaction rates of primary nucleation, elongation and fragmentation. Such a relationship can be well formulated through following interesting scaling laws for the apparent fiber growth rate and the half-time for fibrillation [Hong13], *i.e.*

$$\kappa_{1/2}^{fib} \equiv \dot{M}(t_{1/2})/m_{tot} \propto \left[\left(k_e^+ m_{tot}\right)^{n-1} k_f^+\right]^{1/n},$$

$$t_{1/2}^{fib} \propto \ln\left[\left(k_f^+\right)^{2/n}\Big/\left(k_n^+ m_{tot}^{n_c-1}\right)\right]\left[\left(k_e^+ m_{tot}\right)^{n-1} k_f^+\right]^{-1/n}, \quad (5)$$

where $t_{1/2}^{fib}$ is determined by $M(t_{1/2}^{fib}) = [M(0) + M(\infty)]/2$.

Towards the process of cell damage, whether similar relations are still valid has not been ascertained. Here we explore this problem by randomly varying the model parameters that we are most interested in. As the processes of amyloid formation and cell damage are closely correlated through the number concentration of oligomers in a very complicated way, we let alone those reaction rate constants which are solely related to the amyloid formation, and focus on the effects of protein concentration $m_{tot}$, the number concentration of oligomers $P_{oli}$, the reaction rate constants for oligomer binding $k_b^+$, self-healing of cell membrane $k_b^-$, normal ion exchange $k_l^n$ and abnormal ion leakage $k_l^a$.

Through extensive numerical computations (please see Fig. S4 in the Supporting Materials for details), following relations for the apparent ion leakage rate and the half-time for ion leakage are found, *i.e.*

$$\kappa_{1/2}^{cell} \equiv \dot{I}_+(t_{1/2}^{cell})/I_+(0) \propto k_l^a \left[\left(k_b^+ P_{oli}\right)^{-1} + c_0\right]^{-1/2},$$

$$t_{1/2}^{cell} \sim \left(\kappa_{1/2}^{cell}\right)^{-1}, \quad (6)$$

in which $I_+(t_{1/2}^{cell}) = [I_+(0) + I_+(\infty)]/2$ and $c_0$ is a fitting parameter. The first formula in Eq. 6 could also be replaced by $\kappa_{1/2}^{cell} \propto k_l^a \left[\left(k_b^+ m_{tot}\right)^{-1} + c_0\right]^{-1/2}$, but the agreement seems to be a bit poorer. Within the parameter space we explored, no apparent dependence of $\kappa_{1/2}^{cell}$ and $t_{1/2}^{cell}$ on $k_b^-$ and $k_l^n$ is observed. Besides, both $\kappa_{1/2}^{cell}$ and $t_{1/2}^{cell}$ are found to be independent of the cell concentration $c_{tot}$, which is consistent with the conclusion of dimensionless analysis. But it is noted that this result is only valid when linear Fick's law is adopted for the rates of normal ion exchange and abnormal ion leakage. In the nonlinear region, $\kappa_{1/2}^{cell}$ and $t_{1/2}^{cell}$ should sensitively depend on the cell concentration $c_{tot}$ in general.

**Discussions and Conclusions**

There are many controversies on the possible mechanisms of cell damage caused by amyloid formation. One of representative examples is which form of protein aggregates is responsible for the cytotoxicity? As we have already mentioned in the introductory section, for quite a long time mature fibrils were widely taken as the major cause, but this view has been largely changed in the last several years and now oligomers are considered as the prime pathogenic factor instead. Here we list four representative mechanisms (see Fig. 5), namely the speed for cell membrane damage (or the speed for changing cells from normal to damaged state in the current model) is directly proportional to (i) the number concentration of oligomers, (ii) the mass concentration of oligomers, (iii) the number concentration of aggregates and (iv) the mass concentration of aggregates. The difference between Mechanism (i) and Mechanism (ii) lays on whether the damage to cell membrane is caused by the two ends (like a pin) or the whole surface (like a sticker) of an oligomer. Mechanisms (iii) and (iv) are similar to the first two, except that both oligomers and fibrils are included. Based on our current study (mainly through dissecting Fig. 3 and Fig. 4, data not shown) and other related experiments (especially those on fiber agitation or sonication), mechanism (iv) is usually excluded (*e.g.* $M$ is not affected, but cytotoxicity has been changed a lot during agitation). Furthermore, since oligomer size is generally quite small, it is hard to tell Mechanism (i) from Mechanism (ii) by analyzing the results of mathematical models (this difficulty will be left to experiments and simulations). Finally, in the current model we adopt Mechanism (i) and apply it with great success, but this does not necessarily mean the possibility of Mechanism (iii) for some other amyloid proteins has been excluded. Actually both mechanisms (i) and (iii) can be easily included into a unified model, in which $k_b^+ P_{oli}$ or $k_b^+ P$ is replaced by a comprehensive binding rate function $\sum_{i=n_c}^{\infty} k_b^+(i)[A_i]$. Related studies are left to the future.

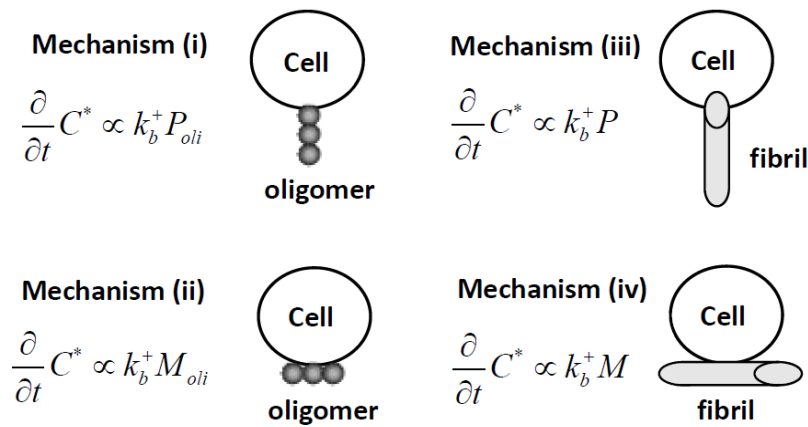

Fig 5. An illustration of four possible mechanisms for cell membrane damage.

In the main text, a two-state model based on whether cell damaged or not is constructed for simplicity. In reality, to account for the degrees of how cells are damaged, a number of states are generally required in order to achieve a satisfactory description. However this problem can be easily solved if a new notation $C(s,w,t)$ is introduced instead of $C(w,t)$ and $C^*(w,t)$, where an additional index $s \in [0,1]$ is added to characterize the different states of a cell based on its damage condition (for example, $s=0$ stands for the normal state $C$, while $s=1$ stands for the totally damaged state $C^*$). Then in a similar way, corresponding 2+1D convection equations can be derived in replace of the last two formulas in Eq. S1. Interested readers can look for details in Supporting Materials.

In this paper, the moment-closure method based on the maximum entropy principle plays a key role in model reduction. Especially the entropy function which we adopted in Eq. S5 directly determines the accuracy of the reduced model. In order to make the static values of $P(t)$, $[C_+(t)-C_-(t)]/(2c_{tot})$ and $1-I_+(t)/c_{tot}$ closer to their exact solutions than what have been shown in Figs. 2A-2D, a more accurate entropy function should be considered. Some initial explorations have been carried out in our previous study [Tan13], but further investigations are still needed. Another issue is related to the self-closure of last three formulas in Eq. 4, which totally lays on the particular choice of linear Fickian law in Eq. 3 as we have shown in Supporting Materials. Once the nonlinearity of ion exchange and ion leakage are considered, high-order moments will be involved automatically. In that case, additional moment-closure method must be introduced as what we have done for the mass-action equations. Related studies are still going on.

As a conclusion, in this paper we have studied the problem of amyloid formation and correlated cell damage by constructing a mathematical model, which consists of infinite ODEs in the form of mass-action equations together with two reaction-convection PDEs. This model is based on four simple assumptions, one of which is that cell damage is raised by oligomers rather than mature fibrils. Then we simplify the model to a system of 5 ODEs by using the maximum entropy principle. With the simplified model, the effects of nucleation and elongation, fragmentation, protein and seeds concentrations on amyloid formation and cell damage are extensively explored and compared with experiments. We hope that our results can provide a valuable insight into the processes of amyloid formation and quantitative roles of oligomers played during cell damage, which is a prerequisite for understanding the happening, proceeding and suspending of amyloid diseases.

**Acknowledgement**

This work was supported by the National Natural Science Foundation of China (Grants 11204150 and 11471185) and by the Tsinghua University Initiative Scientific Research Program (Grants 20121087902 and 20131089184). L. H. also acknowledges the Erasmus Mundus Master Programme in System Biology from Aalto University, Finland.

# Supporting Materials

**Liu Hong\*, Ya-Jing Huang, Wen-An Yong**

Zhou-Pei Yuan Center for Applied Mathematics, Tsinghua University, Peking, 100084, P.R. China

\*zcamhl@tsinghua.edu.cn

**The kinetics of amyloid formation and cell damage can be quantified through coupled mass-action equations and reaction-convection equations.**

Mathematically, the model described in the main context (Eqs. 1-3) can be formulated through the laws of mass-action and basic principles of continuum mechanics , *i.e.*

$$\begin{cases} \dfrac{d[A_1]}{dt} = -n_c k_n^+ [A_1]^{n_c} + n_c k_n^- [A_{n_c}] - 2k_e^+ [A_1] \sum_{j=n_c}^{\infty} [A_j] + 2k_e^- \sum_{j=n_c+1}^{\infty} [A_j], \\ \dfrac{d[A_i]}{dt} = 2k_e^+ [A_1]([A_{i-1}] - [A_i]) - 2k_e^- ([A_i] - [A_{i+1}]) + 2\sum_{j=n_c+i}^{\infty} k_f^+(i, j-i)[A_j] \\ \quad - \sum_{j=n_c}^{i-n_c} k_f^+(j, i-j)[A_i] - 2\sum_{j=n_c}^{\infty} k_f^-(i,j)[A_i][A_j] + \sum_{j=n_c}^{i-n_c} k_f^-(j, i-j)[A_j][A_{i-j}] \\ \quad + (k_n^+ [A_1]^{n_c} - k_n^- [A_i] - 2k_e^+ [A_1][A_{i-1}] + 2k_e^- [A_i])\delta_{i,n_c}, \quad i \geq n_c \\ \dfrac{\partial}{\partial t} C(w,t) = -k_b^+ P_{oli} C(w,t) + k_b^- C^*(w,t) - \dfrac{\partial}{\partial w}\left[k_l^n(w) C(w,t)\right] \\ \dfrac{\partial}{\partial t} C^*(w,t) = +k_b^+ P_{oli} C(w,t) - k_b^- C^*(w,t) + \dfrac{\partial}{\partial w}\left[k_l^a(w) C^*(w,t)\right], \quad w \in [0,1] \end{cases} \quad \text{(S1)}$$

where $[A_1]$, $[A_i]$, $C(w,t)$ and $C^*(w,t)$ stand for concentrations of monomer, oligomers or fibrils of size $i$, normal and damaged cells with ion concentration $w$ at time $t$ respectively. $P_{oli} = \sum_{i=n_c}^{n_o} [A_i]$ is the number concentration of oligomers. The reaction rate constants for fibril fragmentation and association are generally length-dependent. Here we refer to Hill's model [Hill83],

$$\begin{cases} k_f^+(i,j) = k_f^+ (ij)^{n-1}(i\ln j + j\ln i)/(i+j)^{n+1}, \\ k_f^-(i,j) = k_f^- (i\ln j + j\ln i)/[ij(i+j)], \end{cases} \quad \text{(S2)}$$

where $n \approx 1-3$ represents degrees of motion freedom of a fiber in the solution. The initial and boundary conditions of above equations can be generally taken as $[A_1](0) = m_{tot}$, $[A_i](0) = 0$ for $i \geq n_c$, $C(w,0) = c_{tot}\delta(w-1)$, $C^*(w,0) = 0$ for $w \in [0,1]$ and $C(0,t) = C^*(1,t) = 0$ for $t \geq 0$.

Based on the last two PDE in Eq. S1 and boundary conditions, we can easily derive that

$$\begin{cases} \int_0^1 \left[C(w,t) + C^*(w,t)\right] dw = c_{tot}, \\ \dfrac{d}{dt}\int_0^1 w \cdot \left[C(w,t) + C^*(w,t)\right] dw = \int_0^1 \left[k_l^n(w)C(w,t) - k_l^a(w)C^*(w,t)\right] dw. \end{cases} \quad \text{(S3)}$$

The first formula represents the conservation law of total cells, while the second one shows that the total ion changes come from two contributions: the normal ion exchange and abnormal ion leakage as we supposed.

Eq. S1 provides a basic mathematical model for the kinetics of amyloid formation and cell damage. But in practice the consumption of huge CPU time for each computation constitutes a major bottleneck, which therefore calls for a reliable simplification method to speed up the calculation.

**Derivation of moment-closure equations from mass-action equations by the maximum entropy principle.**

Based on the mass-action equations given in Eq. S1, the time-evolution equations for the number concentration of aggregates (including both oligomers and fibrils) $P = \sum_{i=n_c}^{\infty} [A_i]$ and the mass concentration of aggregates $M = \sum_{i=n_c}^{\infty} i \cdot [A_i]$ obey the following form

$$\begin{cases} \dfrac{d}{dt} P = k_n^+ (m_{tot} - M)^{n_c} - k_n^- [A_{n_c}] + \sum_{i=n_c}^{\infty} \sum_{j=i+n_c}^{\infty} k_f^+ (i, j-i) [A_j] \\ \qquad - \sum_{i=n_c}^{\infty} \sum_{j=n_c}^{\infty} k_f^- (i, j) [A_i][A_j], \\ \dfrac{d}{dt} M = n_c k_n^+ (m_{tot} - M)^{n_c} + 2 k_e^+ (m_{tot} - M) P - 2 k_e^- P - (n_c k_n^- - 2 k_e^-)[A_{n_c}], \end{cases} \qquad (S4)$$

which clearly is not closed, as the unknown variables $[A_i]$ are not expressed through $P$ and $M$.

To solve this problem, a systematic moment-closure method, which is based on the principle of maximum entropy, has been proposed and applied with great success in our previous studies [Hong13]. Namely, we seek the solution of the following constrained optimization problem

$$\max \quad S([A_i]) = -k_B \sum_{i=1}^{\infty} ([A_i] \ln [A_i] - [A_i]) \qquad (S5)$$

$$\text{s.t.} \quad \sum_{i=n_c}^{\infty} [A_i] = P, \sum_{i=n_c}^{\infty} i \cdot [A_i] = M, [A_1] + \sum_{i=n_c}^{\infty} i \cdot [A_i] = m_{tot}. \qquad (S6)$$

It could also be translated into an equivalent variational problem

$$\frac{\delta}{\delta [A_i]} \left[ S([A_i])/k_B + \lambda_1 \left( \sum_{i=n_c}^{\infty} [A_i] - P \right) + \lambda_2 \left( \sum_{i=n_c}^{\infty} i \cdot [A_i] - M \right) \right. \\ \left. + \lambda_3 \left( [A_1] + \sum_{i=n_c}^{\infty} i \cdot [A_i] - m_{tot} \right) \right] = 0, \qquad (S7)$$

where $\lambda_1$, $\lambda_2$ and $\lambda_3$ are Lagrangian multipliers. The solution of Eq. S7 is given by

$$\begin{cases} [A_1] = \exp(\lambda_3) \\ [A_i] = \exp[\lambda_1 + i(\lambda_2 + \lambda_3)], \qquad (i \geq n_c) \end{cases} \qquad (S8)$$

based on which $\lambda_1$, $\lambda_2$ and $\lambda_3$ are related to $P$, $M$ and $m_{tot}$ as

$$\begin{cases} \lambda_1 = \ln\left[\dfrac{P^2}{M-(n_c-1)P}\right] - n_c \cdot \ln\left[\dfrac{M-n_c P}{M-(n_c-1)P}\right], \\ \lambda_2 = \ln\left[\dfrac{M-n_c P}{M-(n_c-1)P}\right] - \ln(m_{tot}-M), \\ \lambda_3 = \ln(m_{tot}-M). \end{cases} \quad (S9)$$

Put these formulas back into Eq. S4, we obtain the desired moment-closure equations for amyloid formation in replace of mass-action equations.

**Derivation of moment-closure equations from two reaction-convection equations.**

In the former section, we show how to simplify the mass-action equation for amyloid formation by the method of maximum entropy principle. Now we want to further simplify the remaining two reaction-convection equations into several ODEs. A basic motivation lays in the fact that $C(w,t)$ and $C^*(w,t)$ contain too much detailed information that we generally do not care about; only the percentage of normal and damaged cells and the amount of ions leaked through membrane are widely interested and actually measured in experiments.

First of all, we define the following quantities

$$\begin{cases} C_+ = \int_0^1 \left[C(w,t)+C^*(w,t)\right] dw, \\ C_- = \int_0^1 \left[C(w,t)-C^*(w,t)\right] dw, \\ I_+ = \int_0^1 w\left[C(w,t)+C^*(w,t)\right] dw, \\ I_- = \int_0^1 w\left[C(w,t)-C^*(w,t)\right] dw, \end{cases} \quad (S10)$$

which correspond to first four independent moments of $C(w,t)$ and $C^*(w,t)$. It is noted that $C_+(t)=c_{tot}$ is a constant representing the conservation of total cells. Once the rate constants for normal ion exchange and abnormal ion leakage obey Fick's law, that is $k_l^n(w)=2k_l^n\cdot(1-w)$ and $k_l^a(w)=2k_l^a\cdot w$, we can easily find that

$$\begin{cases} \dfrac{d}{dt}C_- = -k_b^+ P_{oli}(C_+ + C_-) + k_b^-(C_+ - C_-), \\ \dfrac{d}{dt}I_+ = k_l^n(C_+ + C_-) - k_l^n(I_+ + I_-) - k_l^a(I_+ - I_-), \\ \dfrac{d}{dt}I_- = k_l^n(C_+ + C_-) - \left(k_l^n + k_b^+ P_{oli}\right)(I_+ + I_-) + \left(k_l^a + k_b^-\right)(I_+ - I_-), \end{cases} \quad (S11)$$

which constitute a system of self-closed ODEs once $P_{oli}$ is given. However if the nonlinearity of ion exchange and leakage are considered, high-order moments will be involved in Eq. S11. Then additional moment-closure method must be introduced just as what we did in the last section.

**Amyloid formation and cell damage are positively correlated with primary nucleation.**

In this section, we study the effect of primary nucleation on amyloid formation and cell damage. For this purpose, both reactions rate constants for fiber fragmentation and association $k_f^+$ and $k_f^-$ are set to be zero in order to avoid the influence of secondary nucleation. While keeping other model parameters the same as in Fig. 2, the forward and backward reaction rate constants for primary nucleation $k_n^+$ and $k_n^-$ are varied in the same way, *e.g.* tenfold and one tenth of base values, so that the static state will not be affected. It is clearly seen from Fig. S1 that the amount of $M$, $P$, $P_{oli}$ and ion leakage all grows simultaneously with the increase of primary nucleation rates. This result means that by speeding up the primary nucleation, we will accelerate the formation of aggregates (including both oligomers and fibrils in contrast to elongation in the next section) and also the speed of cell damage. Namely, amyloid formation and cell damage are positively correlated with primary nucleation.

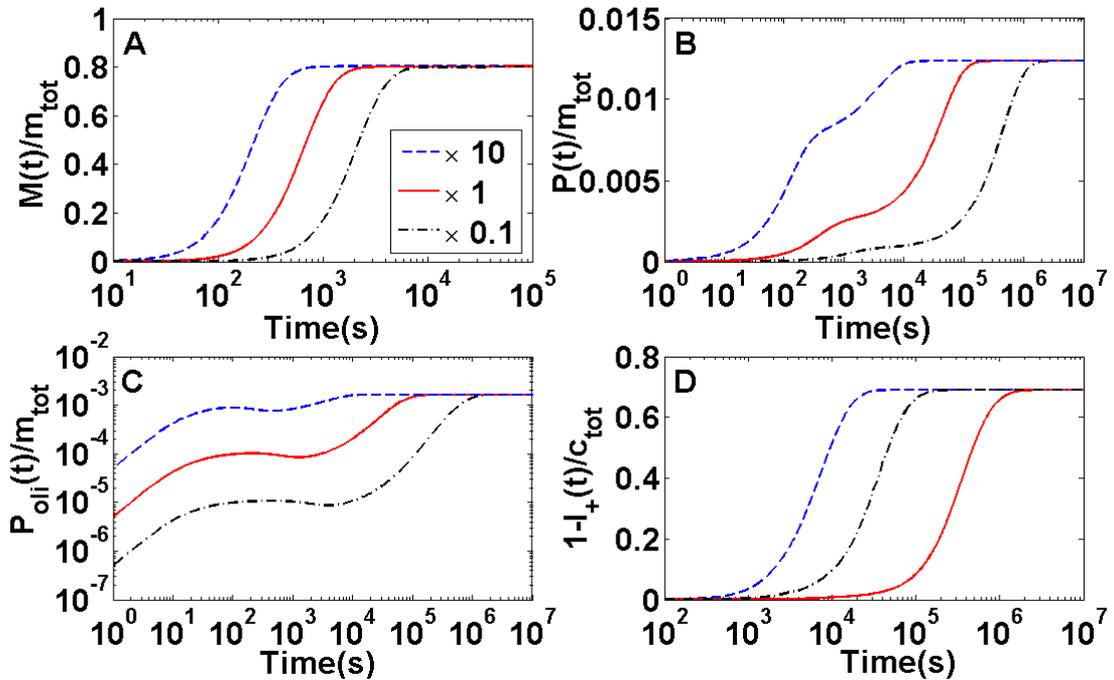

Fig S1. Effects of primary nucleation on amyloid formation and ion leakage. (A) The mass concentration of aggregates; (B) the number concentration of aggregates; (C) the number concentration of oligomers; (D) the amount of ion leakage through cell membrane. Here we set the parameters for red solid lines (base values) as $m_{tot}=5\times10^{-5}M$, $c_{tot}=4.3\times10^{-5}M$, $k_n^+=0.1M^{-1}s^{-1}$, $k_n^-=0.001s^{-1}$, $k_e^+=10^4M^{-1}s^{-1}$, $k_e^-=0.1s^{-1}$, $k_f^+=k_f^-=0s^{-1}$, $k_b^+=10^4M^{-1}s^{-1}$, $k_b^-=0.001M^{-1}s^{-1}$, $k_l^n=5\times10^{-5}M^{-1}s^{-1}$, $k_l^a=1.5\times10^{-4}M^{-1}s^{-1}$, $n_c=2$, $n_o=10$ and $n=1$. The blue dashed lines and gray dotted lines representing two cases with the forward and backward reaction rate constants for primary nucleation varied as tenfold and one tenth of their base values.

**High elongation rate promotes fibril formation, but suppresses oligomer formation and thus ion leakage to some extent.**

Now we try to explore the effect of elongation on amyloid formation and ion leakage. Here we still set $k_f^+ = k_f^- = 0$. And only the forward and backward reaction rate constants for elongation are varied in the same way. Curves shown in Figs. S2A-C tell us that the increase of elongation rate will indeed speed up the formation of fibrils but at the price of less formed oligomers and fibrils. This means that elongation accelerates the kinetics of fibril formation through a fast consumption of nucleuses, in contrast to primary nucleation and secondary nucleation (like fragmentation in the next section) which reach a similar result by generating nucleuses quickly. Since more oligomers will be consumed within a given time (to be exact they grow into fibrils through elongation) and according to our second assumption that oligomers rather than fibrils are harmful to cells, elongation will suppress the degrees of ion leakage and cell damage to some extent (see Fig. S2D). And with a continuous increase of $k_e^+$ and $k_e^-$, the effect of elongation will become less and less apparent due to the limited number of nucleuses.

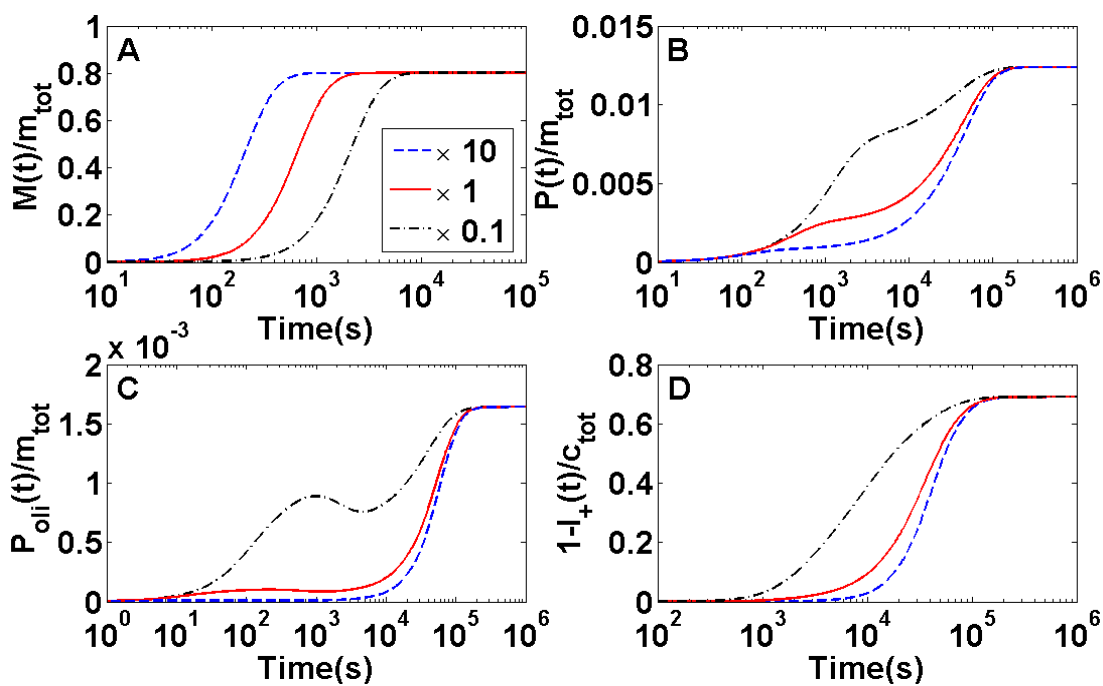

Fig S2. Effects of elongation on amyloid formation and ion leakage. (A) The mass concentration of aggregates; (B) the number concentration of aggregates; (C) the number concentration of oligomers; (D) the amount of ion leakage through cell membrane. Here we set the parameters for red solid lines (base values) as $m_{tot} = 5 \times 10^{-5} M$, $c_{tot} = 4.3 \times 10^{-5} M$, $k_n^+ = 0.1 M^{-1} s^{-1}$, $k_n^- = 0.001 s^{-1}$, $k_e^+ = 10^4 M^{-1} s^{-1}$, $k_e^- = 0.1 s^{-1}$, $k_f^+ = k_f^- = 0 s^{-1}$, $k_b^+ = 10^4 M^{-1} s^{-1}$, $k_b^- = 0.001 M^{-1} s^{-1}$, $k_l^n = 5 \times 10^{-5} M^{-1} s^{-1}$, $k_l^a = 1.5 \times 10^{-4} M^{-1} s^{-1}$, $n_c = 2$, $n_o = 10$ and $n = 1$. The blue dashed lines and gray dotted lines representing two cases with the forward and backward reaction rate constants for elongation varied as tenfold and one tenth of their base values.

**Large fragmentation rate means short fibrils and more oligomers, and thus high cell toxicity.**

In Fig S3, fiber fragmentation is considered and its effect on amyloid formation and cell damage are examined. All the reaction rate constants are taken the same as in Fig. 2, except the forward and backward reaction rate constants for fiber fragmentation and association are changed. Fig. S3A shows that the mass concentration of aggregates is almost unaffected by fragmentation; while the number concentration of aggregates and especially oligomers are both grow quickly at a high fragmentation rate through Figs. S3B-C. So is the speed for ion leakage as a direct consequence (see Fig. S3D). These results confirm the fact that fragmentation will generate more oligomers and short fibrils, and thus high cell toxicity.

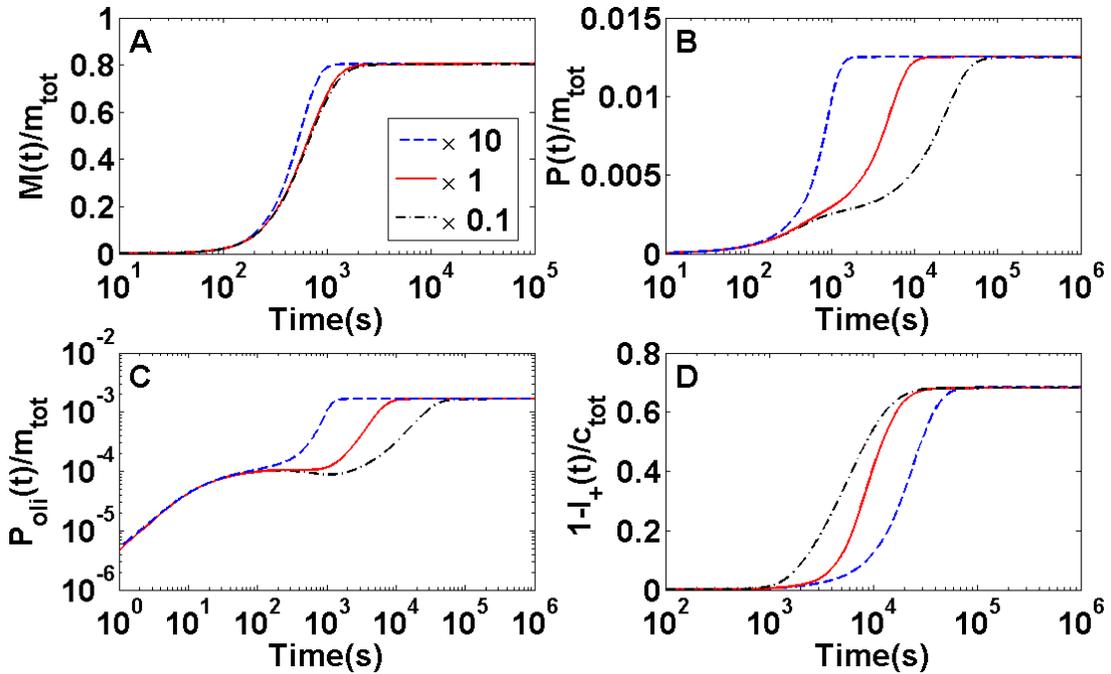

Fig S3. Effects of fragmentation on amyloid formation and ion leakage. (A) The mass concentration of aggregates; (B) the number concentration of aggregates; (C) the number concentration of oligomers; (D) the amount of membrane ion leakage through cell membrane. Here we set the parameters for red solid lines (base values) as $m_{tot} = 5 \times 10^{-5} M$, $c_{tot} = 4.3 \times 10^{-5} M$, $k_n^+ = 0.1 M^{-1} s^{-1}$, $k_n^- = 0.001 s^{-1}$, $k_e^+ = 10^4 M^{-1} s^{-1}$, $k_e^- = 0.1 s^{-1}$, $k_f^+ = 10^{-4} s^{-1}$, $k_f^- = 8 \times 10^4 M^{-1} s^{-1}$, $k_b^+ = 10^4 M^{-1} s^{-1}$, $k_b^- = 0.001 M^{-1} s^{-1}$, $k_l^n = 5 \times 10^{-5} M^{-1} s^{-1}$, $k_l^a = 1.5 \times 10^{-4} M^{-1} s^{-1}$, $n_c = 2$, $n_o = 10$ and $n = 1$. The blue dashed lines and gray dotted lines representing two cases with the forward and backward reaction rate constants for fiber fragmentation varied as tenfold and one tenth of their base values.

**Multiple states of cells can be accounted by a two-dimensional reaction-convection equation.**

In the main text, a two-state model based on whether cell damaged or not is considered for simplicity. However in reality, to account for the degrees of how cells are damaged, a number of states are generally required for a satisfactory description. For this purpose, here we generalize our notation as $C(s,w,t)$ instead of $C(w,t)$ and $C^*(w,t)$, where an additional index $s \in [0,1]$ is introduced to characterize the different states of a cell based on its damage condition (for example, $s=0$ stands for the normal state $C$, while $s=1$ stands for the totally damaged state $C^*$). In a similar way, following 2+1D convection equations can be constructed in replace of the last two formulas in Eq. S1, *i.e.*

$$\frac{\partial}{\partial t}C(s,w,t) = -\frac{\partial}{\partial s}\left[k_b^+(s)P_{oli}C(s,w,t) - k_b^-(s)C(s,w,t)\right] \quad \text{(S12)}$$
$$-\frac{\partial}{\partial w}\left[k_l^n(s,w)C(s,w,t) - k_l^a(s,w)C(s,w,t)\right].$$

Here $C(s,w,t)$ stand for the concentration of cells in the state $s$, with ion concentration $w$ and at time $t$ respectively. $k_b^+(s)$, $k_b^-(s)$, $k_l^n(s,w)$ and $k_l^a(s,w)$ are the generalized rate constants for oligomer binding, cell membrane self-healing, normal ion exchange and abnormal ion leakage respectively.

Similarly, to have the balance laws

$$\begin{cases} \int_0^1\int_0^1 C(s,w,t)dwds = c_{tot}, \\ \frac{d}{dt}\int_0^1\int_0^1 wC(s,w,t)dwds = \int_0^1\int_0^1\left[k_l^n(s,w)C(s,w,t) - k_l^a(s,w)C(s,w,t)\right]dwds, \end{cases} \quad \text{(S13)}$$

we impose following boundary conditions

$$\begin{cases} k_l^n(s,0)C(s,0,t) - k_l^a(s,0)C(s,0,t) = 0, \\ k_l^n(s,1)C(s,1,t) - k_l^a(s,1)C(s,1,t) = 0, \\ k_b^+(0)P_{oli}C(0,w,t) - k_b^-(0)C(0,w,t) = k_b^+(1)P_{oli}C(1,w,t) - k_b^-(1)C(1,w,t). \end{cases} \quad \text{(S14)}$$

**The steady state.**

In the previous study [Hong13], we have shown that the number concentration and mass concentration of aggregates can be approximated as $P(\infty) \sim \left(k_f^+ m_{tot}^{n-1}/k_f^-\right)^{1/n}$ and $M(\infty) \sim m_{tot} - k_e^-/k_e^+$ in the steady state (This result is still valid in the current case since the consumption of oligomers during their interaction with cell membrane is neglected in our model). Based on this result, the steady number and mass concentrations of oligomers could also be obtained as $P_{oli}(\infty) = \sum_{j=n_c}^{n_o}[1-\theta(\infty)]\theta^{j-n_c}(\infty)P(\infty) \approx (n_o - n_c + 1)P^2(\infty)/M(\infty)$ and $M_{oli}(\infty) = \sum_{j=n_c}^{n_o} j\cdot[1-\theta(\infty)]\theta^{j-n_c}(\infty)P(\infty) \approx (n_o + n_c)(n_o - n_c + 1)P^2(\infty)/[2M(\infty)]$, where $\theta(\infty) \approx 1 - P(\infty)/M(\infty)$. In addition, as the cell concentration and ion concentration in the steady state are given as

$$\begin{cases} C_+(\infty) = c_{tot}, \\ C_-(\infty) = \dfrac{c_{tot}\left[k_b^- - k_b^+ P_{oli}(\infty)\right]}{k_b^- + k_b^+ P_{oli}}, \\ I_+(\infty) = \dfrac{c_{tot}k_b^- k_l^n \left[k_b^- + 2k_l^a + k_b^+ P_{oli}(\infty)\right]}{\left[k_b^- + k_b^+ P_{oli}(\infty)\right]\left[k_b^- k_l^n + 2k_l^a k_l^n + k_b^+ k_l^a P_{oli}(\infty)\right]}, \\ I_-(\infty) = \dfrac{c_{tot}k_b^- k_l^n \left[k_b^- + 2k_l^a - k_b^+ P_{oli}(\infty)\right]}{(k_b^- + k_b^+ P_{oli})\left[k_b^- k_l^n + 2k_l^a k_l^n + k_b^+ k_l^a P_{oli}(\infty)\right]}, \end{cases} \quad (S15)$$

we can simplify the ion concentration inside the cells as $I_+(\infty) \sim c_{tot}k_l^n k_b^- / \left[k_l^a k_b^+ P_{oli}(\infty)\right]$, provided that $k_b^+ P_{oli}(\infty) \gg k_b^-$ and $k_l^n \sim k_l^a$, a condition generally holds when the cell damage reaches an apparent status. In this way, the ion concentration inside the cells is found inversely proportional to the number concentration of oligomers $I_+(\infty) \propto P_{oli}^{-1}(\infty)$ in the steady state.

**Kinetic relations in the form of scaling laws.**

In this section, we are going to check the scaling laws held between the kinetic quantities of cell damage and various model parameters, except those which are solely related to amyloid formation.

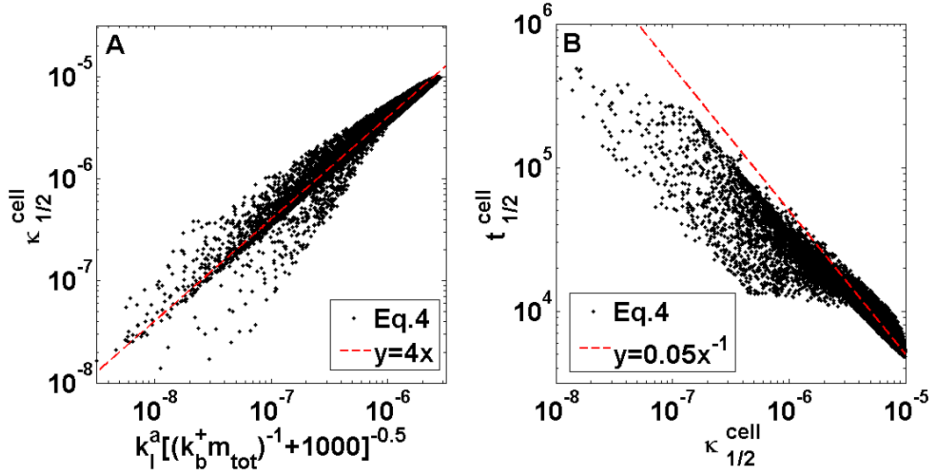

Fig S4. (A-B) Scaling relations among apparent ion leakage rate, the half-time and five model parameters for cell damage. 10,000 data points are generated separately with five model parameters randomly chosen as $m_{tot} = 10^{-8} - 10^{-4} M$, $k_b^+ = 1 - 10^4 M^{-1}s^{-1}$, $k_b^- = 10^{-8} - 10^{-4} M^{-1}s^{-1}$, $k_l^n = 10^{-8} - 10^{-4} M^{-1}s^{-1}$ and $k_l^a = 10^{-8} - 10^{-4} M^{-1}s^{-1}$, while other parameters are fixed the same as in Fig. 2.

**References**
[Hill83] Hill TL. Length dependence of rate constants for end-to-end association and dissociation of equilibrium linear aggregates. *Biophys. J.*, **44**: 285–288, 1983.
[Hong13] Hong L, Yong WA. Simple moment-closure model for the self-assembly of breakable amyloid filaments. *Biophys. J.*, **104**: 533-540, 2013.